\documentclass[nonacm, sigplan]{acmart}

\usepackage{tikz}
\usepackage{amsmath,bm}
\usepackage{subfiles}
\usepackage{balance}
\usepackage{multicol}
\usepackage{framed}
\usepackage{kantlipsum}
\usepackage{lipsum}
\usepackage{setspace}
\usepackage{amsfonts}
\usepackage{enumitem,kantlipsum}
\usepackage[english]{babel}
\usepackage{mdframed}
\usepackage{lipsum}
\usepackage{xcolor}
\usepackage{makecell}
\usepackage{multirow}
\usepackage{float}
\usepackage{algorithm}
\usepackage{algpseudocode}
\usepackage{subcaption}
\usepackage{array}
\newcolumntype{?}{!{\vrule width 1pt}}
\usepackage{booktabs,caption}
\usepackage[flushleft]{threeparttable}
\usepackage{bm}
\usepackage{url}
\usepackage{xfp}
\usepackage{xparse}
\usepackage{expl3}
\usepackage{placeins}
\usepackage[textwidth=0.5in,textsize=tiny]{todonotes}
\usepackage[]{hyperref}

\floatname{algorithm}{Algorithm}
\usepackage[export]{adjustbox}
\usepackage{bbding}

\usepackage{microtype}
\usepackage{cleveref} 



\copyrightyear{2026}
\acmYear{2026}
\setcopyright{acmlicensed}
\acmConference[Conference ASPLOS '26]{Make sure to enter the correct
  conference title from your rights confirmation email}{March 22-26, 2026}{Pittsburgh, USA}
\acmDOI{XXXXXXX.XXXXXXX}

\begin{document}

\title[GoCkpt]{GoCkpt: Gradient-Assisted Multi-Step overlapped Checkpointing for Efficient LLM Training}

\author{Keyao Zhang}
\affiliation{%
  \institution{Zhejiang University \& Alibaba Group}
  \city{Hangzhou}
  \country{China}
}
\email{zhangkeyao@zju.edu.cn}

\author{Yiquan Chen}
\affiliation{%
  \institution{Alibaba Group}
  \city{Hangzhou}
  \country{China}}
\email{zy.zhengyi@alibaba-inc.com}

\author{Zhuo Hu}
\affiliation{%
  \institution{Zhejiang University}
  \city{Hangzhou}
  \country{China}}
\email{zjullk@zju.edu.cn}

\author{Wenhai Lin}
\affiliation{%
  \institution{Zhejiang University}
  \city{Hangzhou}
  \country{China}}
\email{linwh@zju.edu.cn}

\author{Jiexiong Xu}
\affiliation{%
  \institution{Zhejiang University}
  \city{Hangzhou}
  \country{China}}
\email{jasonxu@zju.edu.cn}


\author{Wenzhi Chen}
\authornote{Wenzhi Chen is the corresponding author.}
\affiliation{%
  \institution{Zhejiang University}
  \city{Hangzhou}
  \country{China}}
\email{chenwz@zju.edu.cn}

\renewcommand{\shortauthors}{Keyao Zhang et al.}


\ExplSyntaxOn
\NewDocumentCommand{\Percent}{ m m O{1} } 
{
    \fpeval{ round( (#1) / (#2) * 100, #3) } 
}
\NewDocumentCommand{\PercentChangeA}{ m m O{1} } 
{
    \fp_compare:nNnTF { #2 } = { 0 }
        { \text{未定义} } 
        { 
            \fpeval{ round( ( (#1) - (#2) ) / (#1) * 100, #3) } 
        }
}
\NewDocumentCommand{\PercentChangeB}{ m m O{1} } 
{
    \fp_compare:nNnTF { #2 } = { 0 }
        { \text{未定义} } 
        { 
            \fpeval{ round( ( (#1) - (#2) ) / (#2) * 100, #3) } 
        }
}

\ExplSyntaxOff

\newcommand{\red}[1]{\textcolor{red}{#1}}
\newcommand{\blue}[1]{\textcolor{blue}{#1}}

\begin{abstract}

The accuracy of large language models (LLMs) improves with increasing model size, but increasing model complexity also poses significant challenges to training stability. Periodic checkpointing is a key mechanism for fault recovery and is widely used in LLM training. However, traditional checkpointing strategies often pause or delay GPU computation during checkpoint saving for checkpoint GPU-CPU transfer, resulting in significant training interruptions and reduced training throughput.

To address this issue, we propose GoCkpt, a method to overlap checkpoint saving with multiple training steps and restore the final checkpoint on the CPU. We transfer the checkpoint across multiple steps, each step transfers part of the checkpoint state, and we transfer some of the gradient data used for parameter updates. After the transfer is complete, each partial checkpoint state is updated to a consistent version on the CPU, thus avoiding the checkpoint state inconsistency problem caused by transferring checkpoints across multiple steps. Furthermore, we introduce a transfer optimization strategy to maximize GPU-CPU bandwidth utilization and SSD persistence throughput. This dual optimization—overlapping saves across steps and maximizing I/O efficiency—significantly reduces invalid training time. Experimental results show that GoCkpt can increase training throughput by up to 38.4\% compared to traditional asynchronous checkpoint solutions in the industry. We also find that GoCkpt can reduce training interruption time by 86.7\% compared to the state-of-the-art checkpoint transfer methods, which results in a 4.8\% throughput improvement.
\end{abstract}

\maketitle 

\section{Introduction}
\label{Introduction}

Large language models (LLMs) have undergone rapid adoption across diverse domains in data centers \cite{huCharacterizationLargeLanguage, DBLP:conf/cvpr/HeZRS16}, driven by two key trends: the exponential growth of model parameters and training token volumes. These large autoregressive AI models have demonstrated exceptional performance across a broad range of tasks \cite{bangVTrainSimulationFramework2024}. This trajectory is reinforced by scaling laws, which have propelled model sizes from hundreds of millions of parameters in early models, such as BERT \cite{devlinBERTPretrainingDeep2019}, to over 500 billion in systems like Megatron-Turing NLG \cite{shoeybiMegatronLMTrainingMultiBillion2020}. In this context, mixed-precision training has emerged as the de facto standard for scaling large models. Supported by a robust technological ecosystem—including hardware adaptations (e.g., Tensor Cores), algorithmic innovations (e.g., Automatic Mixed Precision (AMP) \cite{DBLP:conf/dac/GuanHSHW024, DBLP:conf/ppopp/XuSZLH024} with gradient scaling), and framework integration (e.g., PyTorch, TensorFlow)—it delivers measurable efficiency gains (e.g., 1.2-2x training speedups) and resource optimizations (e.g., halving GPU memory footprints for model parameters). By reducing forward propagation precision from FP32 to FP16 or BF16, mixed-precision training significantly lowers training memory demands.

However, the rapid growth in model size has led to drastically extended training durations. Prominent case studies illustrate this trend: the BLOOM model \cite{workshopBLOOM176BParameterOpenAccess2023} required ~3.5 months (~1 million training hours) to train; LLaMA \cite{touvronLlama2Open2023} took 54 days; and OPT-175B \cite{zhangOPTOpenPretrained2022} (175B parameters) trained for ~33 days across 992 NVIDIA A100 GPUs (80GB each) using Fully Sharded Data Parallelism (FSDP) \cite{zhaoPyTorchFSDPExperiences2023a} and 8-way Megatron-LM tensor parallelism to process 180B tokens. Critically, this training endured 105+ restarts due to frequent hardware failures (e.g., overheating, power outages) and software issues (e.g., MPI errors, checkpoint corruption), with the longest stable training interval lasting just 2.8 days. Similarly, GLM-130B \cite{zengGLM130BOpenBilingual2023} (130B parameters) required 60 days of training across 96 DGX-A100 nodes (768 GPUs total).

Large model training faces exceptionally high interruption risks. For example, LLaMA3 experienced interruptions averaging every 3 hours over its 54-day training cycle. Reports from Alibaba's Unicron system indicate a 43.4\% failure rate for resource-intensive LLM training jobs, with 37\% attributed to hardware failures and ~73\% of total failures remediable via restarts \cite{heUnicronEconomizingSelfHealing2023}. Meta's research reveals that 50\% of machine learning training job runtime is squandered due to inefficiencies \cite{hsiaCrossStackWorkloadCharacterization2020}, while Microsoft observes an average failure interval of just 45 minutes in multi-tenant GPU clusters \cite{jeonAnalysisLargeScaleMultiTenant}. Even under nominal operation, training trajectories often exhibit anomalies—slow convergence, stalls, persistent incorrect feature learning, or severe loss fluctuations—that degrade efficiency. Such pathologies, documented in PaLM \cite{chowdheryPaLMScalingLanguage2023} and GLM-130B, and observed recurrently in mainstream models (e.g., BLOOM-175B, OPT-175B), are unpredictable and lack effective prevention strategies. Current mitigations primarily involve rolling back to the latest valid checkpoint and applying corrective actions, such as skipping problematic data batches, adjusting parameter precision, or modifying architectural components.

To mitigate resource waste from interruptions, periodic checkpointing has become a cornerstone technique in large-scale training: PaLM employed a multi-layered strategy (10-minute memory snapshots with full parameter saves at specific intervals), while GLM-130B introduced dynamic interval adjustment. These designs reduced resource loss from hardware failures by 20-30\%, enhancing overall utilization. Indeed, periodic checkpointing is explicitly cited as critical for training continuity in logs of BLOOM-175B and OPT-175B, validating its efficacy in distributed scenarios. In contrast, aperiodic mechanisms (e.g., Just-In-Time (JIT) \cite{guptaJustInTimeCheckpointingLow2024}) reduce overhead via node-level data-parallel replicas but lack generality across diverse distributed setups because some research suggests only using data-parallel inter-node \cite {fanDAPPLEPipelinedData2021, shoeybiMegatronLMTrainingMultiBillion2020}.

However, existing checkpointing mechanisms still face significant efficiency problems. First, computation interrupts remain a critical issue: traditional checkpointing halts or slows down GPU computation during state transfer (especially model/optimizer parameters from GPU to CPU), resulting in low GPU utilization. For instance, saving full model states can take seconds to minutes, leaving GPU compute units idle and directly reducing throughput. Second, synchronization overhead fragments parallelism: existing frameworks \cite{mohanCheckFreqFrequentFineGrained2021, DataStatesLLMLazyAsynchronous} rely on device-level synchronization (e.g., pausing all devices to coordinate writes), breaking the training pipeline and underutilizing GPU-CPU bandwidth. These challenges are exacerbated with cost-effective preemptible instances (60-90\% cheaper than on-demand), where interruptions are frequent—for example, a 64-spot instance cluster experienced 127 training failures within 24 hours \cite{andreExploringLearningRate2022}.

These inefficiencies require a rethinking of the checkpoint architecture. To address computation interrupts and synchronization overhead, we propose GoCkpt, a system that overlaps checkpoint transfers with multi-step training, improving training efficiency through three key innovations.

$\bullet$ Cross-Step Checkpoint Transfer. We propose a scheme that allows checkpoint snapshots to span multiple training steps. This allows training to continue during a checkpoint snapshot, allowing inconsistent checkpoint versions to be transferred to the CPU, reducing training interruptions caused by the checkpoint system and improving training throughput.

$\bullet$ Reconstructing a consistent checkpoint on the CPU. GoCkpt additionally transfers low-precision gradients generated during mixed-precision training steps and updates inconsistent checkpoint versions to the latest consistent version on the CPU. 

$\bullet$ IO bandwidth optimization. GoCkpt introduces I/O optimization techniques to enhance GPU-CPU transmission bandwidth and accelerate data persistence throughput, improving computational resource utilization during training.

We implemented GoCkpt and assessed its performance in comparison to existing systems. The experimental results indicate that this approach surpasses conventional asynchronous checkpointing schemes by 38.4\%. Additionally, compared to state-of-the-art checkpoint transfer methods, GoCkpt reduces training interruption time by up to 86.7\% and enhances throughput by 4.8\%.


The remainder of this paper is organized as follows: \autoref{Background} provides the necessary background; \autoref{Motivation} details our motivation analysis; \autoref{Design and Implementation} describes the design and implementation of GoCkpt; \autoref{Evaluation} presents evaluation results; \autoref{Related Work} discusses related work; and \autoref{Conclusion} concludes.
\section{Background}
\label{Background}

\begin{figure*}[h]
  \centering
  \includegraphics[width=0.9\linewidth]{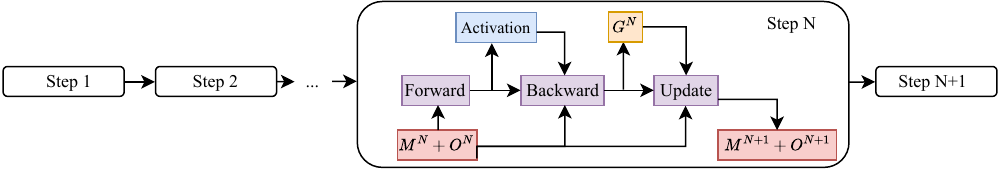}
  \caption{Process of LLM Training (M+O-Model and Optimizer parameters, G-Gradients, N-Model and Optimizer version in step N)}
  \label{fig:process of train}
\end{figure*}

\subsection{Process of LLM Training}

The LLM training process is inherently iterative, involving cyclical interactions between forward propagation, backward propagation, and parameter updates, as schematically depicted in \autoref{fig:process of train}.

At the first part of this workflow lies the Forward pass, where input data traverses the neural network using current model parameters ($M^N$) and optimizer states ($O^N$), producing activations that capture hierarchical semantic representations. Subsequently, the Backward pass computes gradients ($G^N$) of the loss function concerning these parameters, backpropagating errors through the network to quantify sensitivity to input perturbations. Finally, the Update phase leverages these gradients to adjust model weights ($M^N\rightarrow M^{N+1}$) and refine optimizer states ($O^N\rightarrow O^{N+1}$), incorporating techniques such as momentum, gradient clipping, or learning rate scheduling to stabilize training.

A key aspect of this process is versioned state management, where each training step ($N$) is associated with a different snapshot ($M^N+O^N$) of the model and optimizer parameters. This versioning ensures consistency between iterations, enabling incremental progress while reducing the risk of divergence or catastrophic forgetting. Typically, when we checkpoint, we save $M^N$ and $O^N$ as the GPU checkpoint state for the $N$th step.

\subsection{Mixed-Precision Training}
\label{Mixed-Precision Training}

Training billion-parameter models demands not only efficient parallelization but also optimized numerical computation to address memory and speed constraints. Conventional full-precision (FP32) training stores parameters, gradients, and optimizer states (e.g., AdamW's momentum and variance terms) in 32-bit floating-point format, which ensures numerical stability but incurs significant memory overhead—often exceeding GPU/TPU capacity for large models. To address this, mixed-precision training (MPT) has become a cornerstone optimization \cite{liPyTorchDistributedExperiences2020, DBLP:conf/dac/GuanHSHW024, DBLP:conf/ppopp/XuSZLH024}, combining 16-bit (FP16 or Bfloat16) and 32-bit arithmetic to reduce memory usage while preserving model accuracy.

At its core, MPT uses 16-bit arithmetic for computations that tolerate lower precision (e.g., matrix multiplications, forward/backward activation passes) and retains FP32 for operations requiring stability (e.g., gradient updates, optimizer steps). Critically, while optimizer states (e.g., AdamW's m and v) remain in FP32 to avoid instability, a FP32 replica of the model parameters is also kept. The parameters and gradients are stored and computed in 16-bit format. This reduces the memory footprint of parameters and gradients by ~50\% compared to FP32-only training (since FP16 and Bfloat16 use 2 bytes per value vs. 4 bytes for FP32).

Modern frameworks (e.g., PyTorch's torch.cuda.amp, TensorFlow's tf.keras.mixed\_precision) further optimize this by dynamically managing conversions between 16-bit and 32-bit arithmetic during training, ensuring numerical stability (via techniques like loss scaling to prevent gradient underflow) while minimizing overhead. Empirically, MPT has enabled training trillion-parameter models (e.g., GPT-3, PaLM) on thousands of GPUs, where FP32-only approaches would be infeasible due to memory constraints.

\subsection{CPU Assisted Parameters Update}
\label{CPU Assisted Parameters Update}

Some frameworks, such as DeepSpeed Offload++ and Deep Optimizer States \cite{mauryaDeepOptimizerStates2024}, use the CPU to assist the GPU in updating parameters and offloading memory. Similar principles can be effectively applied to optimize checkpointing systems, particularly for state information transmission and storage. By leveraging the CPU, we can pass the latest gradients to older model parameter versions, while the optimizer is consistently updated with the latest GPU-computed values.

\section{Motivation}
\label{Motivation}


Large-scale model training demands extraordinary computational resources, where checkpoint systems play a pivotal role in mitigating the cost of unplanned interruptions. However, existing checkpointing techniques face persistent challenges in GPU utilization, state transfer overhead, and adaptability to dynamic training scenarios. This section first models the time consumption of checkpoint systems to quantify inefficiencies, then analyzes limitations of current solutions, and finally demonstrates how our approach (GoCkpt) leverages mixed-precision training and relaxed consistency constraints to achieve near-zero overhead checkpointing.

\subsection{Quantifying Inefficiencies in Checkpoint Systems}
\label{Quantifying Inefficiencies in Checkpoint Systems}

To systematically evaluate checkpoint overhead, we model the training process as a sequence of steps with periodic checkpointing. Let $T_{\text{total}}$ denote the total effective training time (sum of all step times $T_{\text{step}}$), $N$ the checkpoint interval (number of steps between saves), p the system failure rate (failures per second), $T_{\text{ckpt}}$ the time to save a complete checkpoint, and $T_{\text{load}}$ the time to restore from a checkpoint. 


Training interruptions incur three types of wasted time:  

1. Checkpoint save overhead: Over $T_{\text{total}}$, the total time spent saving checkpoints is $T_{\text{save}}$ = $\frac{T_{\text{total}}}{N \cdot T_{\text{step}}} \cdot T_{\text{ckpt}}$ (since one checkpoint is saved every N steps). 

2. Checkpoint restore induced idle time: Failures can occur uniformly between checkpoints, with an average gap of $\frac{N T_{\text{step}}}{2}$ (assuming exponential distribution for failure times and p is small, a common model for hardware/software errors \cite{benoitCheckpointingYoungDaly2022}). The expected idle time before recovery is $\frac{1}{2} p N T_{\text{total}} T_{\text{step}}$ (by memoryless property of exponential distribution), where $p^{-1}$ is the mean time between failures (MTBF).  

3. Checkpoint load overhead: Each failure requires restoring the latest checkpoint, contributing $T_{\text{restore}} = p T_{\text{total}} T_{\text{load}}$ over $T_{\text{total}}$.  

Total wasted time $T_{\text{waste}}$ combines these components:  

$$T_{\text{waste}} = T_{\text{save}} + \frac{1}{2} p N T_{\text{total}} T_{\text{step}} + p T_{\text{total}} T_{\text{load}}$$
  
(Note: We use $p^{-1}$ instead of $p T_{\text{full}}$ for clarity, aligning with standard reliability theory.)  

The ratio of wasted time to effective training time, $P$ quantifies checkpoint inefficiency:  

$$P = \frac{T_{\text{waste}}}{T_{\text{total}}} = \frac{T_{\text{ckpt}}}{N T_{\text{step}}} + \frac{p N T_{\text{step}}}{2} + p T_{\text{load}}$$

A critical observation is that P is minimized when the derivative with respect to N is zero. Taking $\frac{dP}{dN} = -\frac{T_{\text{ckpt}}}{N^2 T_{\text{step}}} + \frac{p T_{\text{step}}}{2} = 0$. we find the optimal checkpoint interval $N^*$:  

$$ N^* = \sqrt{\frac{2 T_{\text{ckpt}}}{p T_{\text{step}}^2}}$$

So that we can get $P^* = \sqrt{2pT_{\text{ckpt}}}+pT_{\text{load}}$ is the minimized overhead of checkpoint system when $N=N^*$. The corresponding GPU utilization overhead can be calculated by $\frac{P^*}{P^*+1}$.

This confirms our initial intuition: the optimal checkpoint frequency balances save overhead against failure-induced losses. However, our analysis reveals a deeper insight: the fundamental bottleneck of checkpoint systems lies not in the interval N, but in the per-checkpoint overheads $T_{\text{ckpt}}$ and $T_{\text{load}}$.

\subsection{Limitations of Existing Checkpoint Schemes}
\label{problems of existing checkpoint schemes}

Current checkpointing techniques struggle to minimize $T_{\text{ckpt}}$ and $T_{\text{load}}$ for large-scale models. We categorize their limitations as follows:  

Parallelism-constrained schemes: Periodic checkpointers like CheckFreq \cite{mohanCheckFreqFrequentFineGrained2021} and VeloC \cite{nicolaeVeloCHighPerformance2019} attempt to overlap checkpointing with training steps. However, due to strict consistency requirements (e.g., ensuring all GPUs finish the same step before saving), they can only parallelize within a single step’s forward/backward pass, leaving most GPU cycles idle during checkpoint saves. Our experimental validation shows that existing optimizations still offer nearly 5\% throughput improvement at the optimal checkpoint frequency. (\autoref{Evaluation})

Storage-bottleneck solutions: Persistence-focused approaches such as PCCheck \cite{stratiPCcheckPersistentConcurrent2025} and GPM \cite{pandeyGPMLeveragingPersistent2022} use byte-addressable persistent memory (PMEM) or block transfers to accelerate storage. However, these approaches fail to address the GPU-CPU bandwidth bottleneck: transferring checkpoints (e.g., a 16GB checkpoint for a billion-parameter model) over PCIe Gen3 (which can achieve a maximum bandwidth of approximately 12 GB/s) still takes over a second per checkpoint, and even using PMEM cannot reduce this time further. 


\subsection{Feasibility of GoCkpt}
\label{Feasibility of GoCkpt}

To overcome the aforementioned difficulty in reducing GPU interruption duration, a natural approach is to parallelize checkpoint transmission with more training steps, thereby hiding the checkpoint operation under more training computations. However, this presents a serious problem: the checkpoint state changes between training steps. Directly parallelizing checkpoint transmission with training will destroy the consistency of the checkpoint states, resulting in different checkpoint versions for each part and affecting checkpoint accuracy. The characteristics of mixed-precision training and CPU parameter update techniques in \autoref{Mixed-Precision Training} and \autoref{CPU Assisted Parameters Update} inspire us to take advantage of the fact that the gradient space in mixed-precision training is much smaller than the model parameter and optimizer parameter space. In addition to the checkpoint data, we can also transmit a portion of the gradient data generated by backpropagation and use the CPU to update the checkpoint, ultimately developing a complete and consistent checkpoint version on the CPU.

\section{Design and Implementation}
\label{Design and Implementation}


%

%

\begin{figure*}[tbp]
  \centering
  \begin{subfigure}{0.34\linewidth} 
    \centering
    \includegraphics[scale=0.85]{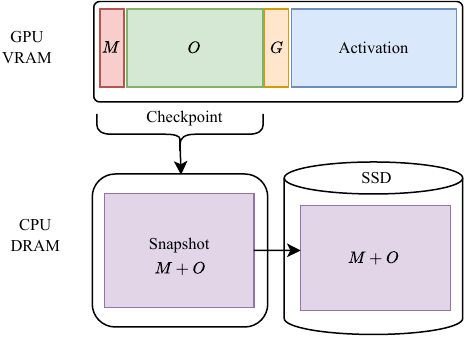}
    \caption{Snapshot in single step}
    \label{fig:single_step}
  \end{subfigure}
  \hfill 
  \begin{subfigure}{0.64\linewidth} 
    \centering
    \includegraphics[scale=0.9]{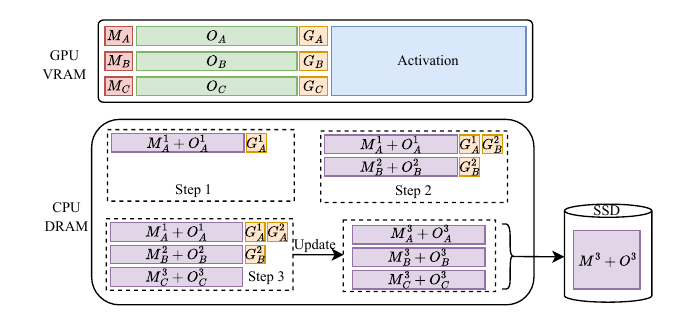}
    \caption{Snapshot in multiple steps}
    \label{fig:multi_step} 
  \end{subfigure}
  
  \caption{Traditional single-step snapshot (a) and GoCkpt multi-step snapshot (b) overview}
  \label{fig:single_multi_step}
\end{figure*}

\begin{figure*}[tbp]
  \centering
  \begin{subfigure}{0.49\linewidth} 
    \centering
    \includegraphics[scale=0.87]{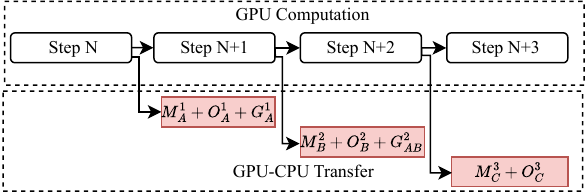}
    \caption{Parallelize GPU-CPU transfer with GPU computation}
    \label{fig:transfer_update}
  \end{subfigure}
  \hfill 
  \begin{subfigure}{0.49\linewidth} 
    \centering
    \includegraphics[scale=0.87]{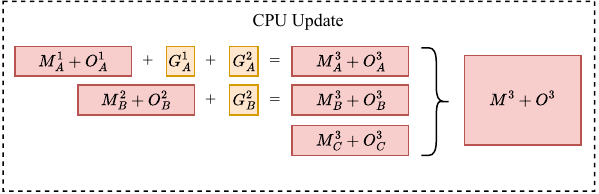}
    \caption{Update the checkpoint state to a consistent version by CPU}
    \label{fig:update_backend} 
  \end{subfigure}
  
  \caption{Compute transfer overlap and CPU-assisted updates}
  \label{fig:Tranfer_and_Update}
\end{figure*}

As discussed in \autoref{Motivation}, existing checkpointing methods share a common limitation: they are unable to mitigate the training latency caused by transferring checkpoint data from the GPU to the CPU.

To minimize this latency, we propose GoCkpt. Our design aims to achieve the following goals:

$\bullet$ Overlap the checkpoint snapshot transfer process with multiple training steps.

$\bullet$ Reconstruct a consistent checkpoint state on the CPU.

$\bullet$ Optimize I/O to fully utilize the PCIe bandwidth between the GPU and CPU and avoid disrupting training.


\subsection{Overview of GoCkpt}
\label{Overview of GoCkpt}

\autoref{fig:single_multi_step} illustrates the overall framework of GoCkpt's design, highlighting the differences between existing single-step checkpoints and our proposed multi-step checkpoints. As \autoref{fig:single_step} shows, traditional checkpointing schemes typically consist of two phases: snapshot and persist. The snapshot phase transfers checkpoint data (primarily model and optimizer states) to the CPU, interrupting training. After the snapshot phase, a background thread is launched to perform the persist phase, persisting the complete checkpoint state on the CPU to a medium such as an SSD. In contrast, as \autoref{fig:multi_step}, our GoCkpt approach consists of three phases: the checkpoint overlap transfer phase, the checkpoint CPU reconstruction phase, and the checkpoint persistence phase. The checkpoint overlap transfer phase also requires transferring checkpoint data to the CPU. Still, by splitting the entire checkpoint data into multiple parts, we can overlap each part with a training step and additionally transfer the low-precision gradient data from mixed-precision training for subsequent phases. The checkpoint CPU reconstruction phase leverages the CPU data to update the checkpoint state to the latest consistent version through CPU computation. The final persistence phase is consistent with traditional checkpointing schemes.

\subsection{Parallelize GPU-CPU transfer with GPU computation}
\label{Parallelize GPU-CPU transfer with GPU computation}

\subsubsection{transfer contens}

GoCkpt hides the visible interruption duration by overlapping the GPU-to-CPU checkpoint transfer with GPU computation. The diagram is shown in \autoref{fig:transfer_update}, where we begin the checkpoint operation after Step N. Instead of transferring all data at once, we split the entire checkpoint into multiple parts, transferring a portion at each step. In the diagram, the checkpoint state, including model and optimizer parameters, is divided into three parts: A, B, and C. These parts overlap with training in Step N+1, N+2, and N+3, with one part transferred in each step. Since each step in large model training updates the model and optimizer parameters based on the gradients generated in that step, the three parts of the checkpoints transferred to the CPU are the checkpoint states corresponding to Steps N+1 through N+3. In addition to the checkpoint state, we also need to transfer the gradients corresponding to the existing checkpoints to the CPU at each step ( $G_A^1$ and $G_{AB}^2$ in the diagram). These gradients are retained only within the corresponding step and therefore need to be transferred to the corresponding step.

\subsubsection{Transmission order and priority management}

Because model parameters and optimizer parameters often reside in separate memory spaces, we organize them into blocks. This ensures that after each block of model parameters is transferred, the corresponding optimizer parameters are immediately transferred, achieving block-level alignment between the two. At the end of each step, we adaptively detect the data blocks that have been assigned and submit the transmission tasks for the gradients corresponding to these data blocks. We use a prioritized queue to manage transmission tasks, with gradient transmission tasks assigned a higher priority. When gradient transmission conflicts with model/optimizer state transmission tasks, gradient transmission takes precedence.

\subsubsection{Checkpoint Stall Analysis}
\label{Checkpoint Stall Analysis}

In this section, we analyze the duration of training interruptions caused by GoCkpt and compare it with traditional asynchronous checkpoints (Async) and the most advanced checkpoint transmission scheme that overlaps with single-step training (Async-O) to demonstrate the theoretical advantages of GoCkpt's transmission scheme.

\begin{figure*}[tbp]
  \centering
  \begin{subfigure}{\textwidth}
    \centering
    \includegraphics[width=\textwidth]{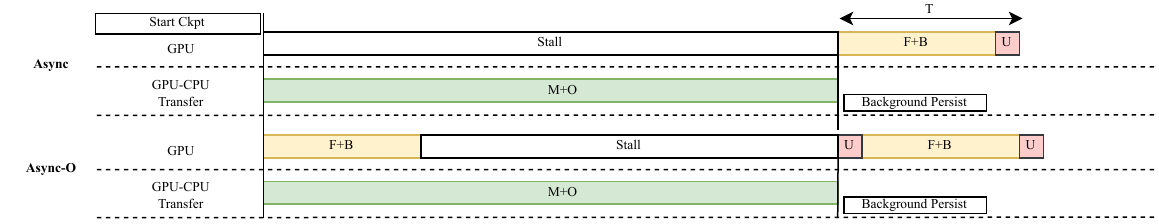}
    \caption{traditional async checkpoint schemes: Async and Async-O}
    \label{fig:Sync-Async Flow}
  \end{subfigure}
  \begin{subfigure}{\textwidth}
    \centering
    \includegraphics[width=\textwidth]{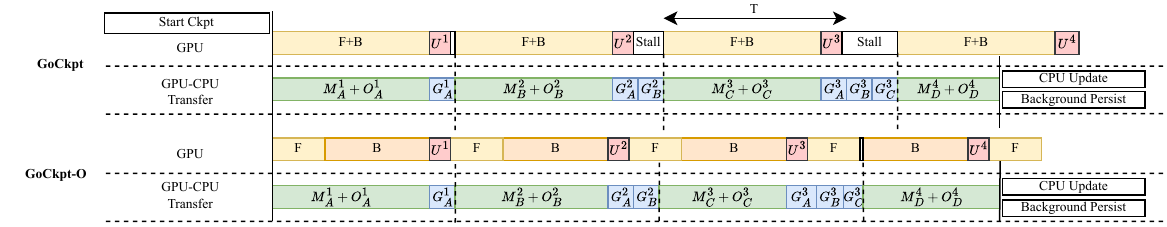}
    \caption{multi-step overlapped checkpoint schemes: GoCkpt and GoCkpt-O}
    \label{fig:GoCkpt Flow}
  \end{subfigure}
  \caption{Computation and data transfer flow of various checkpointing schemes. F-Forward, B-Backward, U-Update, T-Time of one single step, M+O-When GPU transfers Model and Optimizer parameters to CPU, G-When GPU transfers gradients to CPU}
  \label{fig: Computation and data transfer flow}
\end{figure*}

In the actual training of large models, single-step training times can vary slightly due to computational and communication delays. However, this variation is relatively small relative to the total training time and is consistent over time, meaning that the training times for consecutive steps are typically very close. Therefore, we can assume that the single-step training time is always the same and use the average single-step training time of the first few steps during the actual checkpointing process as a reference.
First, we analyze the idea of the GoCkpt algorithm and conduct a detailed analysis of its time cost and the acceleration effect on checkpoints.

As shown in \autoref{fig:Sync-Async Flow}, the typical checkpoint schemes nowadays usually run in parallel with the first GPU-CPU transmission stage after the checkpoint saving process is initiated. The effectiveness of this method depends on the single-step duration: if the single-step duration exceeds the checkpoint transmission time, pauses become inevitable \cite{mohanCheckFreqFrequentFineGrained2021}.

As shown in \autoref{fig:GoCkpt Flow}, our checkpointing system requires transferring the contents of large model checkpoints and the gradient information needed in sections while training steps continue. For the GoCkpt solution, we hide entirely the checkpoint transfer process within the training process, so the only visible overhead to the user is the gradient transfer. In mixed-precision training scenarios, only one-sixth of the gradients need to be transferred for each checkpoint part at each step.

Assuming we split the checkpoint transfer into N steps, we can calculate the checkpoint outage duration for Async-O and GoCkpt separately:

$T_{Async-O}=(N-1)T_{step}$

$T_{GoCkpt}=\sum^{N-1}_{i=1}\frac{i}{7}T_{step}=\frac{N(N-1)}{14}T_{step}$

$\Delta T=T_{GoCkpt}-T_{Async-O}=\frac{-N^2+15N-14}{14}T_{step}$

It can be seen from the formula that, in the optimal case, GoCkpt can overlap the entire checkpoint transmission process into 7 or 8 steps, and can bring a 4$T_{step}$ reduction in checkpoint outage duration. Therefore, we propose an optimal checkpoint transmission strategy: the checkpoint process is overlapped into seven steps. Suppose the model and optimizer state transmission still cannot be completed after seven steps. In that case, the remaining checkpoints are transmitted by blocking, thereby avoiding an increase in gradient transmission overhead that would offset the optimization of the GoCkpt method.

\subsubsection{Further hide the overhead of gradient transmission}
\label{Further hide the overhead of gradient transmission}

Based on the previous scheme, we propose GoCkpt-O, a more optimized scheme to reduce the overhead of gradient transmission further. It is based on some existing implementations—such as DeepSpeed's BF16Optimizer and the Zero-Stage3 implementation that retain gradients from the forward pass of the current step for use in the forward pass of the next step. They do not overwrite the original gradient space with newly computed gradients until the next backpropagation step. This behavior allows us to overlap the transmission of model gradients with the forward propagation of the next step, using gradients that remain unchanged during the parameter update phase of the current step.

By leveraging this mechanism, we can further reduce the communication overhead associated with gradient transfer in the overall training process.

The data transfer flow can be visualized as \autoref{fig:GoCkpt Flow} GoCkpt-O: when the gradient transmission time does not exceed the combined duration of the parameter update and the forward propagation of the next step, the gradient transfer can be effectively overlapped into the forward pass, thereby reducing perceived latency.




\subsection{Update the checkpoint state to a consistent version by CPU}
\label{Update the checkpoint state to a consistent version by CPU}

\subsubsection{Parameter Update Module}
\label{Parameter Update Module}

After transferring the model parameters, optimizer states, and gradients, we update parameters on the CPU using the AdamW optimization strategy. As illustrated in \autoref{fig:update_backend}, consider overlapping checkpoint transfers across three consecutive steps: For checkpoint version 1 of part A transferred at Step N+1, gradients of part A computed in Steps N+1 and N+2 are used to update checkpoint version 3. Similarly, checkpoint version 2 of part B transferred at Step N+2 is updated with part B’s gradients computed at Step N+2, also yielding checkpoint version 3. Once these updates are complete, the CPU retains the full checkpoint version 3—equivalent to directly transferring the checkpoint from Step N+3 to CPU memory.

During the parameter update process, we also use a multi-threading mechanism to update the parameters in parallel, thereby minimizing the overhead of parameter updates. In fact, from the experiments in \autoref{Evaluation}, we can see that the time to update the parameters is relatively short with an appropriate number of threads, much shorter than the interval between checkpoint saves.

\subsubsection{Checkpoint loading}
\label{Checkpoint loading}

Since optimizing checkpoint loading and optimizing checkpoint interruption have orthogonal effects on GPU utilization, we use the same synchronization strategy for loading checkpoints as traditional checkpointing schemes. When loading a checkpoint, it is first read from the SSD into CPU memory and then transferred to GPU memory. Once the model and optimizer parameters are ready on the GPU, training is resumed at the step after the checkpoint is consistent.

\subsection{IO Bandwidth Optimization}
\label{IO Transfer Optimization}

\subsubsection{Multi-threaded Design}

GoCkpt introduces multi-threaded design optimizations in both the snapshot and persistence phases of data transfer. During the multi-step snapshot phase, background threads manage independent CUDA streams and thread synchronization to manage GPU-CPU checkpoint transfers, minimizing disruption to training. During the persistence phase, multiple threads concurrently write to the SSD to maximize NVMe SSD bandwidth utilization.

\subsubsection{GPU-CPU PCIe Transfer Optimization}
To optimize data transfer between the GPU and CPU, we pre-register the CPU memory used as Pinned Memory using the PyTorch interface. This technique locks the memory pages to be used, avoiding swapping in and out and improving data transfer efficiency. We also split the model and optimizer parameters into 4MB chunks, transferring the corresponding model and optimizer parameter chunks at a time. This maximizes PCIe bandwidth while ensuring traceability of transfer status.

\subsubsection{Data persistence module}
\label{Data persistence module}

After the parameter update is complete, the data persistence module uses multiple threads in parallel to save the model parameters to disk in the background, fully utilizing the SSD's write bandwidth. After the model tensors are persisted, a callback function is used to save the pre-prepared checkpoint metadata, such as the current step count, historical throughput, and historical training time. Saving this metadata marks the completion of the latest checkpoint, allowing it to be used for recovery. If the previous checkpoint metadata has not yet been written to disk when the next checkpoint is saved, GoCkpt will wait for the last checkpoint to complete before starting the new checkpoint process.

\subsection{Multi-card Environment Design}
\label{Multi-card Environment Design}

In a multi-GPU environment, training frameworks like DeepSpeed launch a training process per GPU, so the checkpointing framework only needs to handle inter-GPU synchronization (with Rank 0 monitoring completion by other Ranks and coordinating synchronization).

For checkpointing strategies in multi-GPU setups, we define rules based on actual parallelism (tensor parallelism, pipeline parallelism). Each GPU stores only its shards involved in parallel training, focusing on data parallelism efforts.

Under data parallelism, we determine how data parallel group members collectively save exact model copies. With ZeRO-1/2 parallelism, each GPU in the data parallel group saves its optimizer state shard; during loading, these shards are fetched via intra-group communication, avoiding redundant duplication \cite{rajbhandariZeROMemoryOptimizations2020}.

When the model is trainable, data parallelism reduces per-GPU memory pressure. Thus, our approach combines tensor parallelism (TP) on individual GPUs with ZeRO-1 data parallelism across multi-GPU servers to distribute workload.



\subsection{GoCkpt Implementation}
\label{GoCkpt Implementation}

We implemented our GoCkpt and GoCkpt-O solutions in Python and C++, totaling approximately 2,000 lines of code, and integrated them into the Deepspeed training framework. We used Pybind11 to provide easy-to-use interfaces, including three API interfaces: save\_checkpoint, backward\_begin, and update\_begin, which are inserted before the checkpoint save, backpropagation, and parameter update phases of each step. We used PyTorch for CPU-bound memory allocation and management, C++ for multithreading, and independent CUDA stream management for fine-grained, high-performance data replication. Finally, we encapsulated Deepspeed's native CPU AdamW updater to implement parameter updates on the CPU, ensuring consistency and accuracy with those on the GPU.

\section{Evaluation}
\label{Evaluation}

\subsection{Setup}
\label{Setup}

The experimental hardware environment was configured with two distinct setups: a single-GPU platform and a multi-GPU server cluster. For the single-GPU configuration, we deployed a system equipped with a Tesla V100S GPU, paired with dual 48-core CPUs (96 cores total), 128 GB of DDR4 RAM, and 3.8 TB of high-speed NVMe SSD storage dedicated to data I/O. In the multi-GPU setup, an Alibaba Cloud server was utilized, featuring eight H100-80GB GPUs interconnected via NVLink for low-latency communication. Each GPU was assigned a dedicated network interface card (NIC), and the GPUs were logically grouped into NUMA nodes with four GPUs per node. The server was powered by a 224-core Intel Platinum 8480C CPU, supported by 2 TB of ECC RAM (composed of 32 × 64 GB modules) and four 3.5 TB NVMe SSDs (aggregating to 14 TB of raw storage capacity). To mitigate potential SSD bandwidth bottlenecks during checkpointing, we implemented targeted GPU optimizations: four GPUs were strategically distributed across 2 NUMA nodes, with each process allocated 28 CPU cores to enforce strong CPU-GPU affinity within the same NUMA domain. Additionally, each GPU independently saved checkpoint data to its associated NVMe SSD, enabling parallel I/O operations and reducing storage access latency.

\subsection{Benchmark and Baseline}
\label{Benchmark and Baseline}

The experimental evaluation focuses on the Wikidata dataset, primarily quantifying model throughput (samples processed per second). Following large-scale model training standards (per LLaMA documentation), raw data was preprocessed via fixed-sequence padding to ensure uniform input dimensions across runs, minimizing computational/memory load variability for consistent checkpointing scheme comparisons.

Diverse models—LLaMA3.1-1B, Qwen3-0.6B, OPT-350M were selected to assess performance across computational scales to smaller configurations. LLaMA3-8B in multi-GPU settings further enabled exploration of scaling behavior, particularly critical distributed communication/synchronization overheads.

Three checkpointing categories were evaluated: synchronous schemes (e.g., Deepspeed, DCP) pausing training to serialize model states (parameters, optimizers, gradients) for strong consistency/recoverability but significant interruption latency; asynchronous schemes (e.g., Torch-Snapshot, DCP-Async) offloading persisting to background processes for concurrent training; and single-step overlapping schemes (DLRover-Flash, Datastates-LLM) interleaving checkpointing with single-step forward/backward computation to reduce overhead.

We replicated two GPU-CPU transfer-optimized schemes: Async (asynchronous with I/O optimizations) and Async-O (combining transfers with single-step overlap). Proposed schemes—the GoCkpt (explicit gradient transfer waits) and the GoCkpt-O (implicit gradient transfer)—were also evaluated, alongside an ideal zero-overhead scenario (no storage delays/computation interruptions) as a theoretical upper bound.

All experiments used consistent batch sizes to isolate checkpointing impacts: one sample/device for single-GPU runs and four samples/device for multi-GPU setups. This scaling ensured full parallel resource utilization while maintaining cross-configuration comparability, enabling evaluation of each scheme under realistic training conditions.
 
\subsection{Performance Evaluation}
\label{Performance Evaluation}

\begin{figure*}[tbp]
  \centering
    \includegraphics[width=\linewidth]{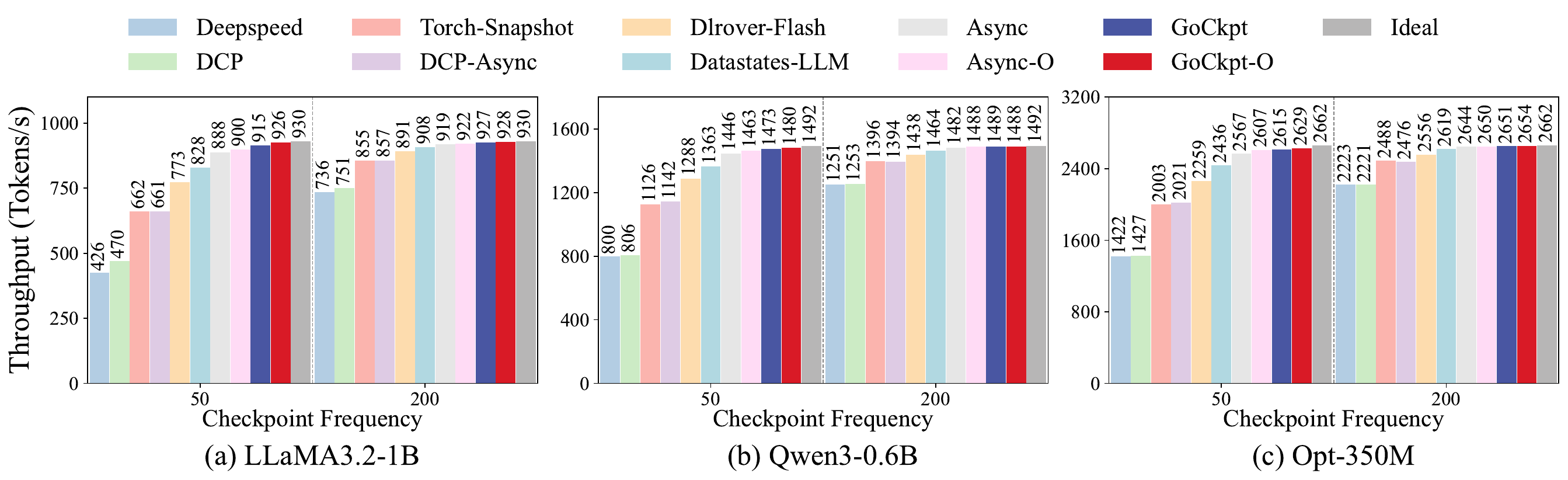}
  
  \caption{Checkpoint throughput and stall time for different checkpoint frequency settings (higher is better)}
  \label{fig:Checkpoint_throughput}
\end{figure*}

In this part, we wish to investigate the role of the GoCkpt checkpointing system in reducing checkpoint downtime and improving the training throughput of large models.

We show the throughput effect of three different models with two different frequencies of checkpoint intervals, once every 50 steps and once every 200 steps.

In \autoref{fig:Checkpoint_throughput}, we can see that under different models and checkpoint frequencies, the gap between GoCkpt and Ideal is between 0.2\% and 1.8\%, and the gap between GoCkpt-O and Ideal is less than 1.2\%. Compared with traditional asynchronous checkpointing schemes, GoCkpt and GoCkpt-O achieve 6.8\% to 38.4\% and 6.7\% to 40.1\% performance improvement, respectively. GoCkpt and GoCkpt-O achieve at most 1.7\% to 3.0\% and 2.9\% to 4.3\% performance improvement, respectively, compared with the two replication schemes Async and Async-O with the same data transfer optimization.

\begin{figure}[tbp]
  \centering
  \includegraphics[width=\linewidth]{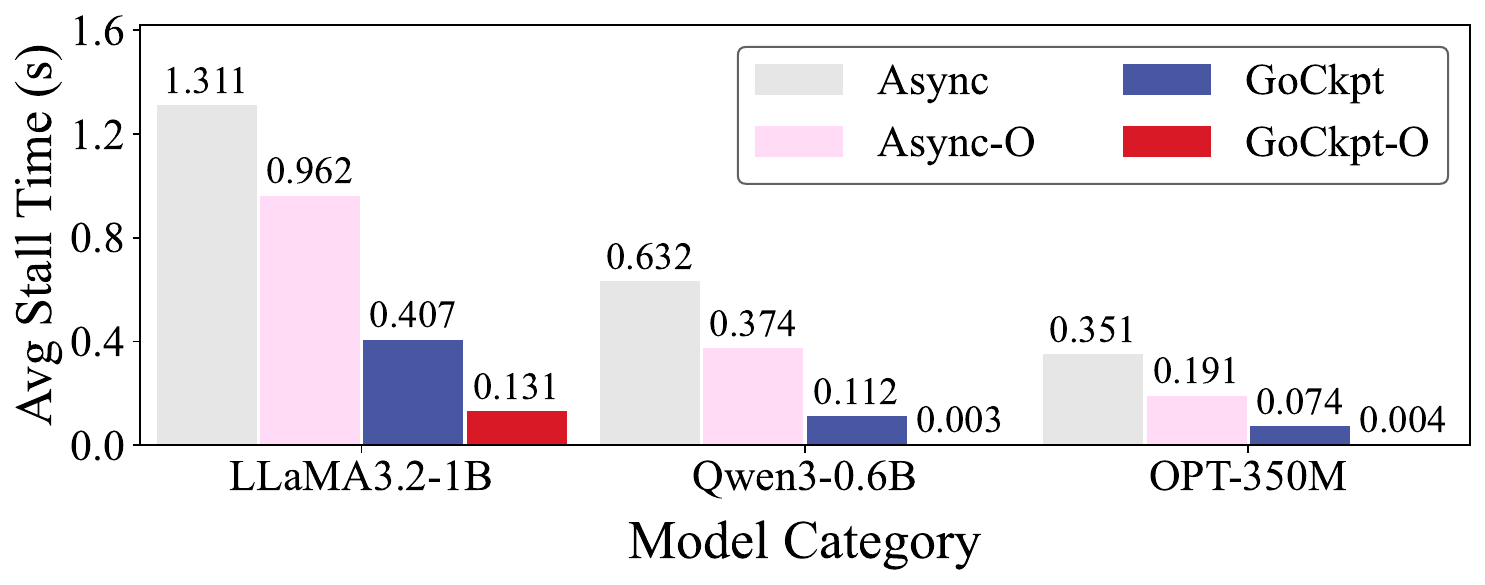}
  \caption{Average stall time of different checkpointing schemes (lower is better)}
  \label{fig:stalltime}
\end{figure}

\autoref{fig:stalltime} shows the comparison of checkpointing time cost between several checkpointing schemes with IO optimization. In the test of LLaMA3.2-1B model, the actual checkpointing time caused by GoCkpt and GoCkpt-O schemes is only 17.7\% to 31.0\% and 0.5\% to 10.0\% of asynchronous checkpointing. Compared with the Async-O scheme, GoCkpt and GoCkpt-O schemes can reduce the checkpointing time by 57.7\% to 70.1\% and 86.4\% to 99.2\%, respectively. For relatively small models such as Qwen3-0.6B and OPT-350M, the GoCkpt-O schemes reduce the time consumption of the checkpointing scheme to 0.003 to 0.004 s, resulting in nearly zero overhead.

\subsection{Checkpoint \& Restore Expriments}
\label{Checkpoint & Restore Expriments}

\begin{table}[]
\begin{tabular}{cccc}
\hline
\textbf{Scheme} & \textbf{Max $T_{ckpt}$(s)} & \textbf{$N_{best}$} & \textbf{Throuput(Tokens/s)} \\ \hline
Deepspeed & 36.79 & 472 & 411.9 \\ \hline    
DCP-Async & 12.226  & 272 & 697.8 \\ \hline
Async     & 1.313  & 89 & 758.0 \\ \hline
Async-O   & 0.988  & 77 & 776.3 \\ \hline
GoCkpt    & 0.435  & 51 & 786.4 \\ \hline
GoCkpt-O  & 0.175  & 32 & 794.1 \\ \hline
\end{tabular}
\caption{The performance of different solutions on the crash expression task, using Llama3.2-1B model, $N_{best}$ is calculated by formula $\sqrt{\frac{2T_{ckpt}}{p T^2_{step}}}$, the system crashes every 600s}
\label{table:crash_expr}
\end{table}

\def\ThputDeepspeed{411.9}
\def\ThputDCPAsync{697.8}
\def\ThputAsync{758.0}
\def\ThputAsyncO{776.3}
\def\ThputGockpt{786.4}
\def\ThputGockptO{794.1}

We simulate the process of crashing the checkpoint system and reloading the latest checkpoint under the simulated failure frequency (crash per 600s), which is used to verify the complete overhead change caused by the checkpoint system.

For the traditional synchronous checkpoint, asynchronous checkpoint, Async, Async-O, and our schemes GoCkpt and GoCkpt-O, we determined the optimal checkpoint frequency according to their checkpoint interruption time, respectively, as shown in \autoref{table:crash_expr}. For each scheme, we picked the checkpoint frequency that is very close to the optimal frequency.

At this frequency, the overhead of checkpoint saving and the total overhead of checkpoint recovery of each kind of checkpoint are close to the lowest in theory. In this case, we verify the optimal throughput of various checkpoint schemes through experiments. From the experimental results, we can see that GoCkpt-O and GoCkpt schemes can effectively improve the optimal checkpoint frequency and obtain higher throughput because they can reduce the time of checkpoint transmission itself. Compared with the existing schemes, GoCkpt-O achieves \Percent{\ThputGockptO}{\ThputDeepspeed}\% and \Percent{\ThputGockptO}{\ThputDCPAsync}\% compared with the synchronous checkpointing scheme Deepspeed and the asynchronous checkpointing scheme torch DCP Async, respectively. Even compared with the transport optimized Async and Async-O schemes, GoCkpt-O can obtain \PercentChangeB{\ThputGockptO}{\ThputAsyncO}\%-\PercentChangeB{\ThputGockptO}{\ThputAsync}\% throughput improvement respectively.


\subsection{Breakdown analysis of GoCkpt system}
\label{Breakdown analysis of GoCkpt system}

\begin{figure*}[tbp]
  \centering
    \begin{subfigure}{\linewidth} 
      \centering
      \includegraphics[width=\linewidth]{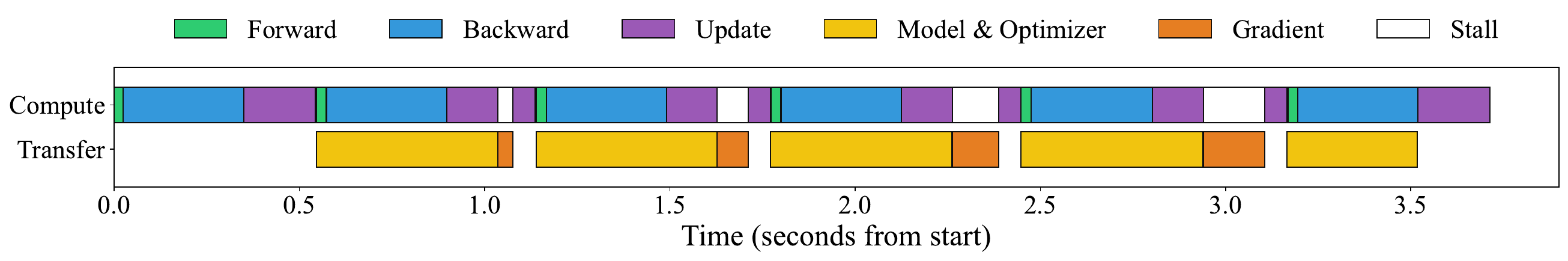}
      \caption{GoCkpt Compute and Transmission breakdown}
      \label{fig:breakdown_grad}
    \end{subfigure}
    \begin{subfigure}{\linewidth} 
      \centering
      \includegraphics[width=\linewidth]{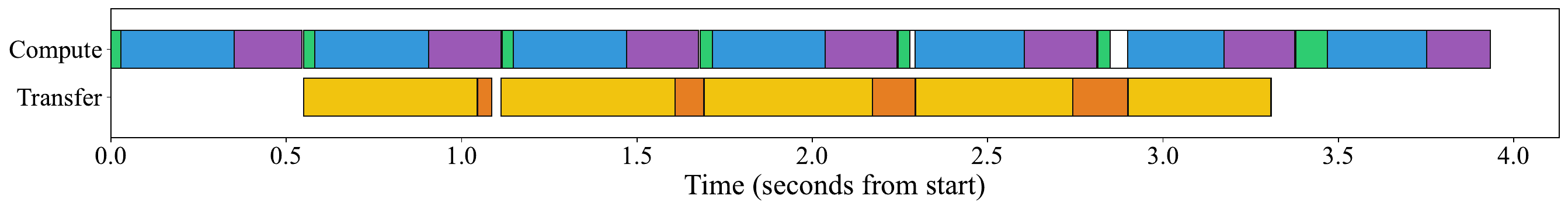}
      \caption{GoCkpt-O Compute and Transmission breakdown}
      \label{fig:breakdown_half_zero}
    \end{subfigure}
    \caption{Breakdown analysis of GoCkpt and GoCkpt-O}
\end{figure*}


In this section, we conduct an experimental analysis of the actual performance of the GoCkpt system under the setting of LLaMA3.2-1B model. We analyze the breakdown of the computation and transmission process during the checkpoint saving process in \autoref{fig:breakdown_grad} and \autoref{fig:breakdown_half_zero}.

As shown in \autoref{fig:breakdown_grad}, we overlap the entire checkpoint transfer process into multiple training steps, and we can see that the actual stall process occurs during the gradient transfer. The figure illustrates the time spent on gradient computation, parameter update, and checkpoint transmission during the checkpoint saving process. The x-axis represents the time in seconds, while the y-axis shows the different phases of the process. The GoCkpt scheme stops training in the process of transferring gradients. Although the training is still paused, the gradient size in the mixed-precision training is much smaller than the size of the model checkpoint, so the majority of the training time can be overlapped with the checkpoint transmission time. In the figure, we can also observe that the update phase is divided into two parts: the part that does not interrupt the checkpoint transmission and the part that does. We analyze that this part of the time without interrupting the checkpoint transmission is an asynchronous operation that is not entirely completed during the backpropagation phase. After the gradient transfer, the parameters are updated on the GPU.

Similarly, we conduct breakdown analysis for the GoCkpt-O scheme in \autoref{fig:breakdown_half_zero}. In this scheme, we assume that the gradient is retained until the end of the forward propagation process, to make full use of the parameter update phase and the overhead of the forward propagation phase to hide the gradient transmission. It can be seen in the real experiments that by applying this optimization, the overhead of gradient transfer is completely hidden in the first few training steps.




\subsection{System Indicator Monitoring}
\label{System Indicator Monitoring}


\begin{figure}[!tbp]
  \centering
  \includegraphics[width=\linewidth]{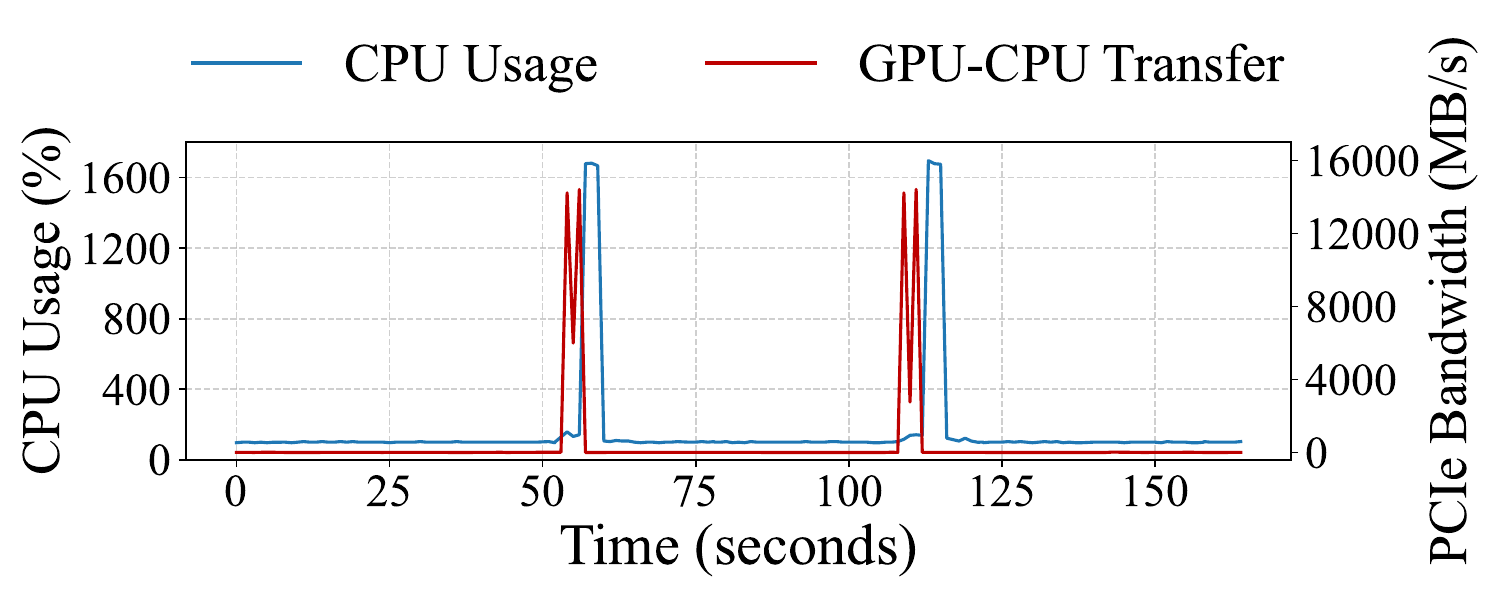}
  \caption{GoCkpt CPU Usage (Left y-axis) and GPU-CPU transfer bandwidth (Right y-axis) curve}
  \label{fig:cpu_usage_grad}
\end{figure}
\begin{figure}[!tbp]
  \centering
  \includegraphics[width=\linewidth]{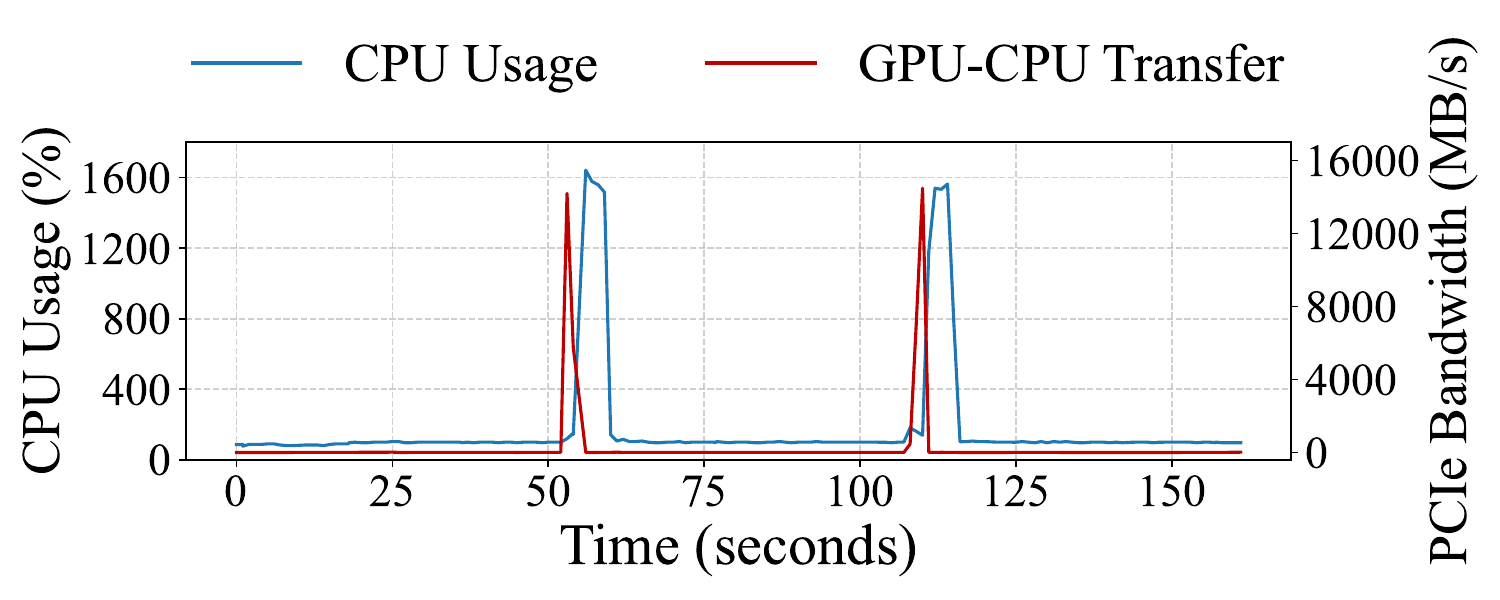}
  \caption{GoCkpt-O CPU Usage (Left y-axis) and GPU-CPU transfer bandwidth (Right y-axis) curve}
  \label{fig:cpu_usage_halfz}
\end{figure}
\begin{figure*}[tbp]
  \centering
  \includegraphics[width=\linewidth]{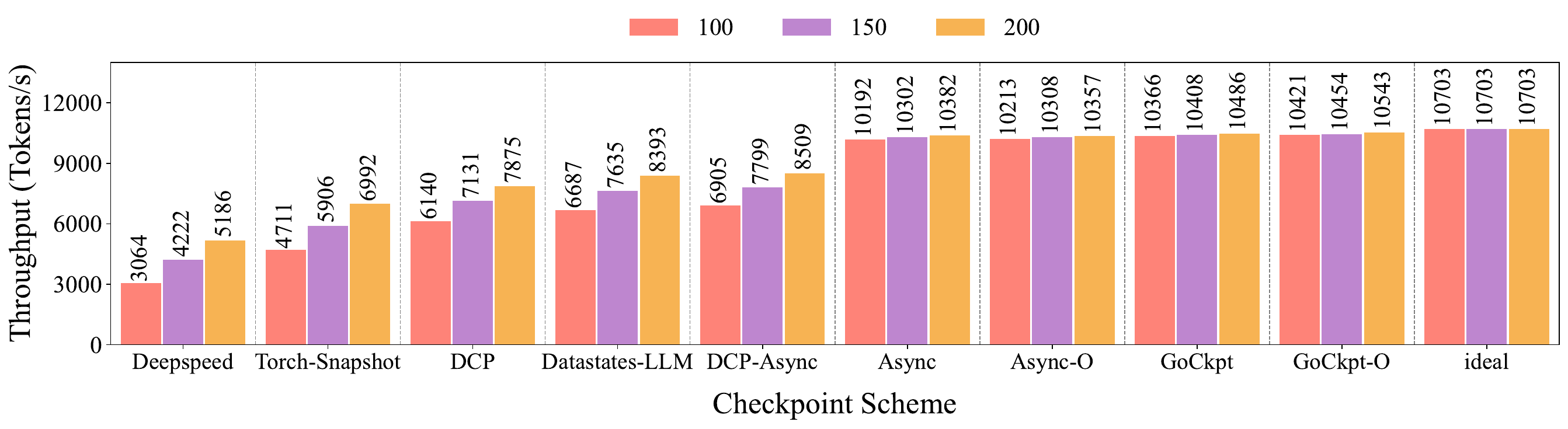}
  \caption{Throughput of LLaMA3-8B with varying checkpoint frequencies (higher is better)}
  \label{fig:multi_card}
\end{figure*}

As can be seen from \autoref{fig:cpu_usage_grad} and \autoref{fig:cpu_usage_halfz}, each checkpoint save operation results in a period of increased GPU-CPU transfer bandwidth and a brief increase in CPU utilization. The actual peak CPU utilization depends on the number of background threads, which is set to 16 in our implementation.

For both GoCkpt and GoCkpt-O, we schedule parameter updates and background persistence after the GPU-CPU transfer. As can be seen in the figure, each GPU-CPU transfer peak is followed by a CPU utilization peak. For both schemes, GoCkpt waits and stalls at each step of the data transfer, reducing the overall bandwidth utilization during the checkpoint transfer. This is reflected in the double peaks in each GPU-CPU transfer in the figure. This means that the maximum bandwidth cannot be achieved throughout the transfer. In GoCkpt-O, by introducing hidden and overlapping gradient transfers, a more balanced single-GPU-CPU transfer is achieved; that is, the data can be viewed as a continuous transfer from the GPU to the CPU. After the transmission is completed, the CPU performs background parameter updates and persistence work to obtain complete checkpoint data from the data transmitted in multiple steps.






\subsection{Scalability Verification Experiments}
\label{Scalability Verification Experiments}

We conducted experiments on a multi-card checkpointing system (4xH100). We found that while multi-card communication occurs via NVLink, the GPU-CPU transmission path in our configuration is undertaken via a PCIe switch, with each card connected to a separate PCIe root domain. This means each card is connected to a separate PCIe switch entity and has its own GPU-CPU transmission path, which does not conflict with the others.

As shown in \autoref{fig:multi_card}, experiments show that when training the LLaMA8B model with four cards, GoCkpt and GoCkpt-O achieve the highest training throughput, trailing Ideal by only 2.0\% to 3.1\% and 1.5\% to 2.6\%, respectively. This represents a 23.2\% to 50.9\% improvement over the asynchronous checkpointing solution DCP-Async, and 1.0\% to 2.2\% throughput improvement over our replicated Async solution.




\section{Related Work}
\label{Related Work}

\textbf{Periodic checkpointing} As a cornerstone technology for large-scale model training, periodic checkpointing has undergone substantial optimization in four core dimensions: \textit{frequency}, \textit{Snapshot}, \textit{Persistence}, and \textit{Loading}. For frequency optimization, studies derive optimal intervals through mathematical modeling (e.g., the Young/Dali formula~\cite{benoitCheckpointingYoungDaly2022}) or graph-based fault correlation analysis~\cite{gholamiestahbanatiMultilevelCheckpointRestart2019}, while JIT~\cite{guptaJustInTimeCheckpointingLow2024} and CPR~\cite{maengCPRUnderstandingImproving} further align with these analyses. Regarding the Snapshot stage, early methods (e.g., CheckFreq~\cite{mohanCheckFreqFrequentFineGrained2021} and VeloC~\cite{nicolaeVeloCHighPerformance2019}) introduced asynchronous schemes to overlap snapshots and training, while later methods (e.g., DataStates-LLM~\cite{DataStatesLLMLazyAsynchronous} and ByteCheckpoint~\cite{wanByteCheckpointUnifiedCheckpointing}) focused on utilizing the GPU-CPU PCIe bandwidth. Innovations in the persistence phase aim to minimize storage latency: PCCheck~\cite{stratiPCcheckPersistentConcurrent2025} overlaps GPU-CPU data transfers with SSD persistence through block-based pipelining; GEMINI~\cite{wangGEMINIFastFailure2023} stores state replicas in network nodes to bypass slow node persistence; GPM~\cite{pandeyGPMLeveragingPersistent2022} explores the feasibility of PMEM as an alternative persistence medium. During the loading phase, ServerlessLLM~\cite{fuServerlessLLMLowLatencyServerless2024} reduces recovery time through in-memory preloading, although this approach relies on error-free local checkpoints, which is a challenge in large-scale training. It is worth noting that some periodic methods relax consistency requirements during persistence (e.g., CPR~\cite{maengCPRUnderstandingImproving}), but are still limited to specific workloads (e.g., recommendation models) and suffer from convergence risks in large language models~\cite{qiaoFaultToleranceIterativeConvergent2019}.

\textbf{Aperiodic Checkpointing} JIT~\cite{guptaJustInTimeCheckpointingLow2024} proposes a fault-triggered checkpointing to avoid periodic overhead, yet still incurs GPU-CPU transfer and persistence costs, making it unsuitable for frequently preempted instances. Swift~\cite{DBLP:journals/tpds/ZhongSLYW24} reconstructs states via pipeline communication logs. FlowCheck\cite{huangFlowCheckDecouplingCheckpointing2025} maintains CPU parameters during training but faces state inconsistency over time, requiring periodic validation. What's more, Phoenixos~\cite{phoenixos} integrates checkpointing at the OS level but lacks application-specific optimizations for large models.

\textbf{Checkpoint Compression} Techniques aim to reduce checkpoint volume: Check-N-Run\cite{eisenmanCheckNRunCheckpointingSystem}, SSDC~\cite{xiangSSDCScalableSparse2024} saves incremental parameters for recommendation models; ExCP~\cite{liExCPExtremeLLM} uses compression/pruning; Inshrinkerator\cite{Inshrinkerator} supports lossy or lossless quantization-aware differential compression; and APR~\cite{akturkACRAmnesicCheckpointing2020} omits recomputable data;Delta-DNN~\cite{huDeltaDNNEfficientlyCompressing2020}, SCAR~\cite{qiaoFaultToleranceIterativeConvergent2019}, LC-Checkpoint~\cite{chenEfficientConstructionsCheckpoints2020b}, QD-Checkpoint~\cite{jinDesignQuantizationBasedDNN2023}, explore lossy compression based on state-difference.




\textbf{Fault Tolerance via Replication} Some works, such as Varuna \cite{athlurVarunaScalableLowcost2022}, Oobleck \cite{ jangOobleckResilientDistributed2023}, and Bamboo \cite{thorpeBambooMakingPreemptible} improve robustness using data parallel replicas, where failed node states are reconstructed from peers. This complements (but is orthogonal to) checkpointing-based strategies.
\section{Conclusion}
\label{Conclusion}

GoCkpt is a method for training large language models (LLMs) that minimizes training interruptions caused by the checkpointing system by overlapping checkpointing with multiple single-step training processes. For large model training, the bottleneck of checkpointing is that the single-step checkpointing interruption time cannot be further reduced, which limits the frequency of checkpointing and leads to significant GPU utilization overhead. Existing checkpointing schemes inevitably incur the overhead of transferring model and optimizer state from the GPU to the CPU. GoCkpt addresses this bottleneck by allowing checkpointing to be overlapped across multiple training steps and restoring a consistent checkpoint version on the CPU. In our experiments, our checkpointing system improves training throughput by 38.4\% compared to traditional asynchronous checkpointing schemes. We also show that GoCkpt can reduce training interruption time by 86.7\% while improving throughput by 4.8\% compared to state-of-the-art checkpointing schemes.


\bibliographystyle{ACM-Reference-Format}
\balance
\bibliography{references}

\end{document}